\documentclass[runningheads,a4paper]{llncs}

\usepackage{amssymb}
\usepackage{graphicx}
\usepackage{enumitem}
\usepackage{courier}

\usepackage{url}
\urldef{\mailsa}\path|{r.k.moore, b.mitchinson}@sheffield.ac.uk|
\newcommand{\keywords}[1]{\par\addvspace\baselineskip
\noindent\keywordname\enspace\ignorespaces#1}

\begin{document}

\mainmatter  

\title{A Biomimetic Vocalisation System for MiRo}

\titlerunning{A Biomimetic Vocalisation System for MiRo}

%
%
\author{Roger K. Moore\inst{1} \and Ben Mitchinson\inst{2}}
\authorrunning{A Biomimetic Vocalisation System for MiRo}

\institute{Dept. Computer Science, University of Sheffield, UK \and Dept. Psychology, University of Sheffield, UK\\
\mailsa}

%
%


\toctitle{A Biomimetic Vocalisation System for MiRo}
\tocauthor{Roger K. Moore, Ben Mitchinson}
\maketitle

\setcounter{footnote}{0}

\begin{abstract}
There is increasing interest in the use of animal-like robots in applications such as companionship and pet therapy.  However, in the majority of cases it is only the robot's physical appearance that mimics a given animal.  In contrast, \emph{MiRo} is the first commercial biomimetic robot to be based on a hardware and software architecture that is modelled on the biological brain.  This paper describes how \emph{MiRo}'s vocalisation system was designed, not using pre-recorded animal sounds, but based on the implementation of a real-time parametric general-purpose mammalian vocal synthesiser tailored to the specific physical characteristics of the robot.  The novel outcome has been the creation of an `appropriate' voice for \emph{MiRo} that is perfectly aligned to the physical and behavioural affordances of the robot, thereby avoiding the `uncanny valley' effect and contributing strongly to the effectiveness of \emph{MiRo} as an interactive device.
\end{abstract}

\section{Introduction}

Recent times have witnessed increasing interest in the use of animal-like robots in applications such as companionship and pet therapy.  For example, \emph{PARO} \cite{paro} is an interactive robotic seal that is particularly popular for therapeutic use in hospitals and care facilities where a live animal would be problematic.  Like \emph{PARO}, the majority of such `zoomorphic' devices are engineered to support specific use-cases, and it is often only the robot's physical appearance that mimics a given animal.  In contrast, \emph{MiRo} \cite{miro} (designed and built by Consequential Robotics Ltd. in collaboration with the University of Sheffield) is the first commercial robot to be controlled by a hardware and software architecture that is specifically modelled on the biological brain \cite{Mitchinson2016,Collins2015a}.

\begin{figure}[ht]
        \centering
        \includegraphics[width=0.8\linewidth]{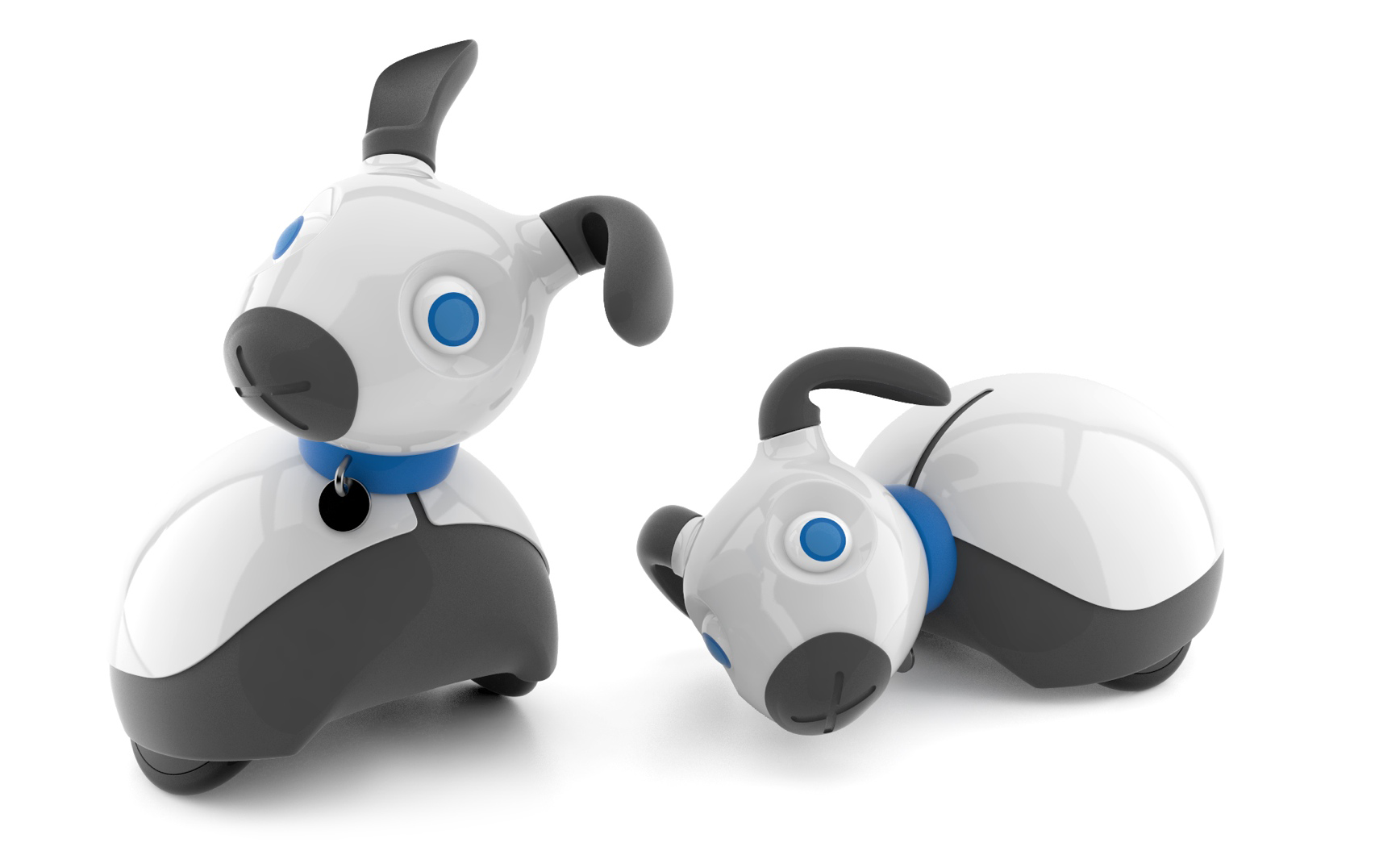}
        \caption{{\it MiRo: the world's first commercial biomimetic robot (available from Consequential Robotics Ltd \cite{miro}).}}
        \label{fig:MIRO}
      \end{figure}

\emph{MiRo} is a highly featured, low-cost, programmable mobile developer platform, with a friendly animal-like appearance, six senses, a nodding and rotating head, moveable hearing-ears, large blinking seeing-eyes, and a responsive wagging tail.  It has been designed to look like a cartoon hybrid of a generic mammal (see Fig.~\ref{fig:MIRO}).  A unique biomimetic control system allows \emph{MiRo} to behave in a life-like way: for example, listening for sounds and looking for movement, then approaching and responding to physical and verbal interactions.

Of special interest here is \emph{MiRo}'s ability to vocalise.  In particular, it was considered important that the vocal output generator should be grounded in a biomimetic model of an appropriate physical sound production apparatus rather than based, for example, on pre-recorded animal sounds.  As a consequence, \emph{MiRo}'s voice was designed using a real-time parametric general-purpose mammalian vocal synthesiser \cite{Moore2016} tailored to the particular physical and behavioural characteristics of the robot.  

This paper presents the biomimetic principles underlying \emph{MiRo}'s vocalisation system, and describes how they have been integrated into the robot's overall architecture.  Section~\ref{sec:MAM} reviews the principles underlying mammalian vocalisation, and Section~\ref{sec:ARCH} outlines \emph{MiRo}'s overall control architecture.  Section~\ref{sec:VOC} then describes how the particular characteristics of \emph{MiRo}'s voice were first designed using a general-purpose mammalian vocal synthesiser, and subsequently implemented on the robot platform itself.  Finally, Section~\ref{sec:CONC} concludes with some observations about the effectiveness of the derived solution.

\section{Vocalisation in Mammals} \label{sec:MAM}

The majority of animals make sound, and different species make sound in different ways.  For example, many insects rub body parts together (a process known as `stridulation'), birds create their songs using a vocal organ known as a `syrinx', and \emph{land}\footnote{Some mammals are adapted for the air or for water.  However, such animals tend to be the extremes in terms of size (for example, the bumblebee bat measures only 30mm, whereas the blue whale is over 30m long) and exploit different mechanisms for generating sound than the majority of land mammals.} mammals typically generate sound using a `larynx' \cite{Hopp1998}.  

\begin{figure}[ht]
        \centering
        \includegraphics[width=\linewidth]{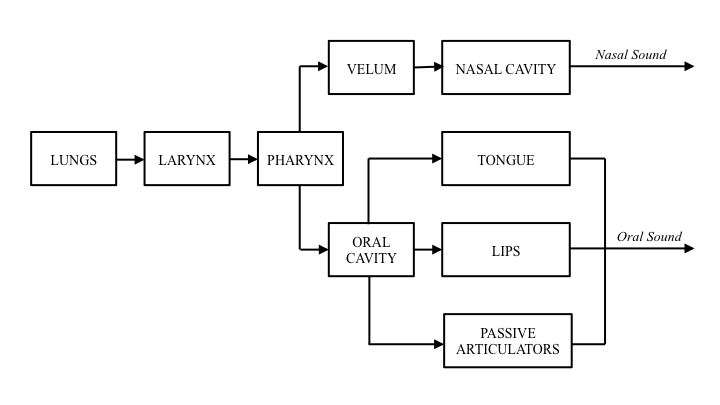}
        \caption{{\it Schematic diagram of the mammalian vocal tract.}}
        \label{fig:VT}
      \end{figure}

The vocal tract for a typical land mammal consists of a larynx, a pharynx, an oral cavity and a nasal cavity (see Fig.~\ref{fig:VT}).  These anatomical features evolved primarily for breathing and eating.  However, over time they have been recruited to create and shape sound. The main sound source is the larynx, which contains a set of `vocal folds' (sometimes referred to as `vocal cords' or the `glottis') - a pair of elastic membranes that are held apart while breathing, but brought together when eating (in order to stop unwanted material from entering the lungs).  The consequence of this arrangement is that air from the lungs that is forced through the closed vocal folds causes them to vibrate.  This creates acoustic energy in the form of a harmonic-rich buzzing sound with a distinct fundamental frequency (perceived as the `pitch' of the voice).  Muscles in the larynx control the length and tension of the vocal folds which, in turn, determine the pitch and timbre of the generated sound.

The rest of the vocal tract can be regarded as a set of interconnected acoustic tubes.  The pharynx lies immediately above the larynx, and this contains the epiglottis: an elastic cartilage that controls entry to the trachea (for breathing) or the oesophagus (for swallowing)\footnote{It is this arrangement that enables mammals to choke!}.  Above the pharynx the acoustic tube splits into two: the main oral cavity (containing the tongue and terminating at the mouth and lips) and the nasal cavity (terminating at the nose).  Airflow into the nasal cavity is controlled by the `velum', a flap-like structure at the back of the mouth.  All of these tubes resonate at different frequencies depending on their size and shape, hence they each filter the spectrum of the harmonic-rich excitation generated by the vibration of the vocal folds; energy is enhanced at some frequencies (the resonances) and suppressed at others.  Vocal tract resonances are known as `formants', and the formants arising from resonances in the oral cavity are of particular significance since they can be altered by moving the tongue and opening/closing the mouth\footnote{In human beings, these are the primary anatomical features used for speaking.}.

In order to determine the characteristics of the sounds produced by a mammalian vocal tract, it is necessary to model ($a$) the `airflow' (the rate at which air is expelled from the lungs), ($b$) the `excitation' (the sound generated by the larynx) and ($c$) the `filtering' (arising from the resonant cavities).  Unsurprisingly, many of these processes are influenced by the size of the animal, since body mass has a direct impact on the physical and acoustic properties of the relevant anatomical components: the lungs, vocal folds, tongue and mouth.

\subsection{Airflow} \label{sec:AFL}

The main source of energy for mammalian vocalisation derives from the lungs, and the key factors are the amount of air that can be stored (`lung capacity') and the rate at which it can be expelled (`airflow').  In general, lung capacity $C$ (in millilitres) scales linearly with body mass $M$ (in kilograms) \cite{Stahl1967} as follows:
\begin{equation} \label{eq:LC}
  C = 53.5  \times M^{1.06} .
\end{equation}

Obviously, airflow is related to breathing, and the breathing rate $B$ (in Hertz) is given by \cite{Worthington1991} as:
\begin{equation} \label{eq:BR}
  B = 0.84 \times M^{-0.26} .
\end{equation}

However, breathing, and hence vocalisation, only uses a proportion of the air in the lungs ($\sim$42\%), and it also restricts airflow (by a factor of 2.62) \cite{Moore2016}.  This means that the volumetric flow rate $Q$ (in litres per second) is given by:
\begin{equation} \label{eq:FR1}
  Q = \frac{0.42 \times C}{2.62 \times \left(\frac{1}{2 \times B}\right)} ,
\end{equation}
which simplifies to:
\begin{equation} \label{eq:FR2}
  Q= 0.32 \times C \times B .
\end{equation}

These parameters characterise the amplitude and duration of each vocalisation and, as can be seen, the predicted value for airflow ($Q$) is directly related to the size of the animal ($M$).  However, these are mean values, and variation around the mean is possible.  For example, a higher airflow would give rise to a shorter but louder vocalisation and \emph{vice versa}.

\subsection{Excitation} \label{sec:EXC}

As air is forced through the closed vocal folds, it escapes in bursts as the folds are momentarily pushed apart.  After each bubble of air escapes, the Bernoulli effect causes the vocal folds to snap shut again, and this action generates a pulse of acoustic energy that propagates through the rest of the vocal tract.  This sequence of events repeats at semi-regular intervals giving rise to a periodic excitation signal with energy at the fundamental frequency of vibration and its associated harmonics\footnote{To a first approximation, the signal generated by the glottis may be modelled as a `sawtooth' waveform.}.

According to \cite{Fletcher2004}, the mean fundamental frequency $F$ (in kHz) of the vocal fold vibration for animals ranging in size from mice to elephants is related to the body mass of the animal by the expression:
\begin{equation} \label{eq:FX}
  F = M^{-0.4} .
\end{equation}
In other words, small animals have high-pitched vocalisations and large animals have low-pitched vocalisations.

The `timbre' of a vocalisation is a function of ($a$) the regularity of the vocal fold vibrations, ($b$) the relationship between the fundamental frequency and its associated harmonics and ($c$) the degree of turbulence in the airflow.  The latter means that,  in addition to fully `voiced' sounds, the mammalian larynx is capable of generating aspirated (breathy) `unvoiced' sounds by holding the vocal folds close together and allowing a small amount of continuous airflow\footnote{Noisy unvoiced excitation at the vocal folds gives rise to whispering in human speech.}.  All of these aspects can be modelled by suitable shaping of the excitation waveform and by the injection of an appropriate level of random noise.

\subsection{Filtering} \label{sec:FLT}

The vocal tract resonances (formants) act as `band-pass' filters which enhance acoustic energy at their resonant frequencies and suppress acoustic energy at other frequencies.  This has a shaping effect on the harmonic-rich spectrum of the excitation signal emanating from the larynx\footnote{It is the different placement of formants that gives rise to the production of different vowel sounds in human speech.}.

The frequencies of the different formants can be estimated by assuming that the vocal tract is a uniform acoustic tube\footnote{This ignores the action of the tongue, which is an appropriate approximation for a non-human land mammal.} which is closed at the vocal folds and open/closed at the mouth.  As the mouth closes, so the formants move down in frequency \cite{Titze2001}.  Hence, the resonant frequency of the $n$th formant $R_n$ (in Hz) can be approximated by the equation:
\begin{equation} \label{eq:FR}
  R_n = (2n-(m+1)) \times \frac{c}{4 \times L} ,
\end{equation}
for $n = 1, 2, 3,:$, where $m$ is the degree of mouth opening (0 $=$ open, 1 $=$ closed), $c$ is the speed of sound (in cm/sec) and $L$ is the length of the vocal tract (in cm)  \cite{Moore2016}.  

According to  \cite{Riede1999}, vocal tract length is correlated with body size:
\begin{equation} \label{eq:VTL}
  L = 3.15 + (11.53 \times \log{M}) .
\end{equation}
This means that large animals have long vocal tracts and thus low formant frequencies (and \emph{vice versa}).  It also means that the distribution of formant frequencies in a vocalisation provides information to a listener about the size of the animal\footnote{Interestingly, animals such as the Red Deer are able to lengthen their vocal tract by lowering their larynx when vocalising, thereby giving the impression of being much larger than they really are \cite{Fitch2001}.}.

\subsection{Summary}

The foregoing provides a complete specification of the minimum set of parameters necessary to simulate the vocalisation of a generic land mammal (as described in \cite{Moore2016}).  The novel contribution here is the mapping of this specification onto \emph{MiRo}'s particular physical characteristics and control architecture.

\section{MiRo's Control Architecture} \label{sec:ARCH}

\begin{figure}[h!]
        \centering
        \includegraphics[width=0.65\linewidth]{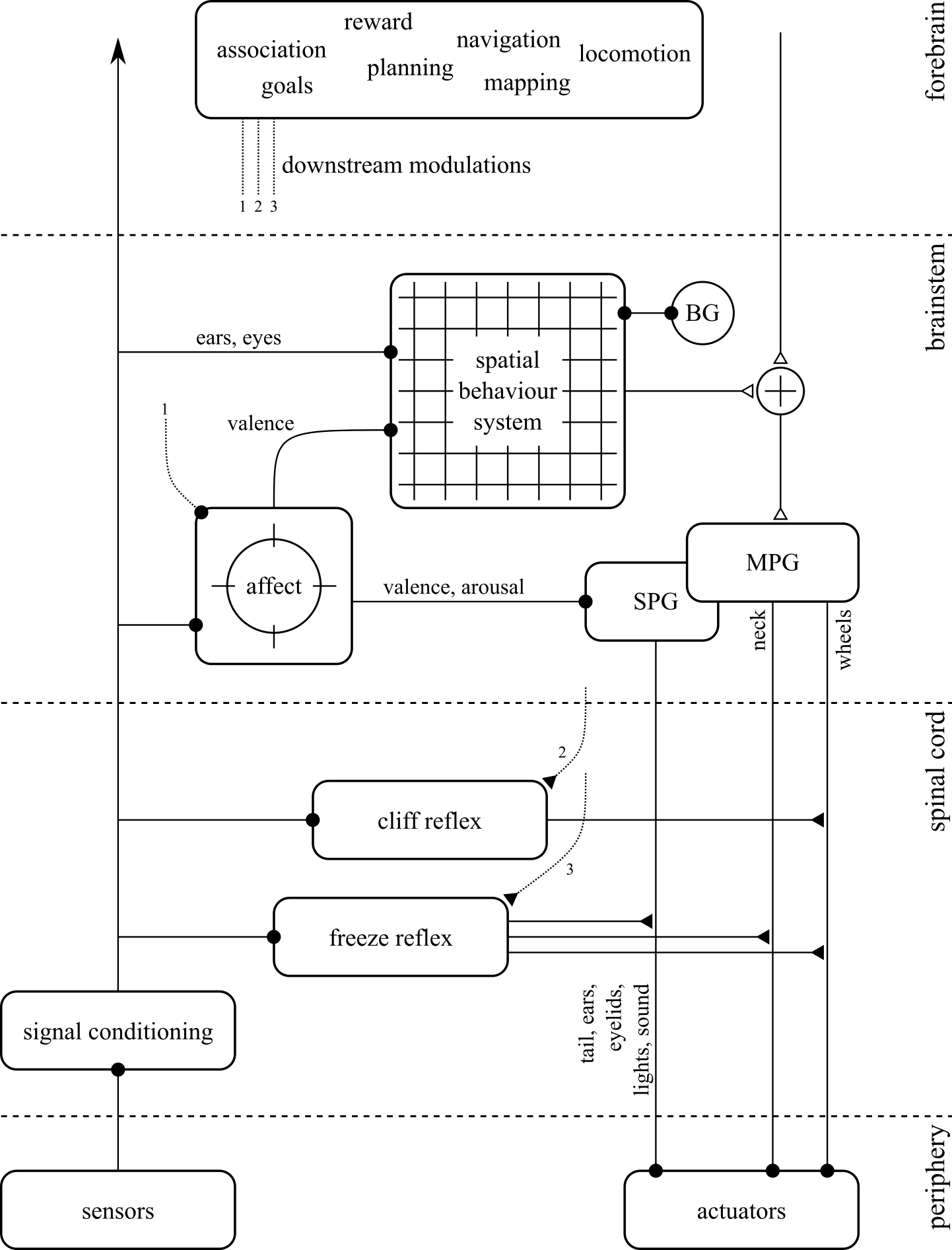}
        \caption{{\it Illustration of MiRo's control architecture loosely mapped onto brain regions (spinal cord, brainstem, forebrain).  Signal pathways are excitatory (open triangles), inhibitory (closed triangles), or complex (closed circles).  BG is the Basal Ganglia.  SPG and MPG are the Social and Motor Pattern Generators.}}
        \label{fig:ARCH}
      \end{figure}

\emph{MiRo}'s control architecture operates across three embedded ARM (Advanced RISC Machines) processors that mimic aspects of spinal cord, brainstem and forebrain functionality (including their relative speed and computational power) - see Fig.~\ref{fig:ARCH}.  One important feature is that the control latency of loops through the lowest reprogrammable processor can be as low as a few milliseconds.  If required, \emph{MiRo} can be operated remotely through WiFi or Bluetooth, and can also be configured as a Robot Operating System (ROS) node \cite{ROS}.

\subsection{Actuators} \label{sec:ACT}

\emph{MiRo} is constructed around a differential drive base and a neck with three Degrees of Freedom (DoF).  Additional DoFs include rotation for each ear, tail droop and wag, and eyelid open/close.  All DoFs are equipped with proprioceptive sensors, and the platform also has an on-board loudspeaker.

\subsection{Sensors}

\emph{MiRo} is equipped with stereo cameras in the eyes, stereo microphones in the base of the ears and a sonar range-finder in the nose.  Four light-level sensors are placed at each corner of the base, and two infrared `cliff' sensors point down from its front face.  Eight capacitive sensors are arrayed along the inside of the body shell and over the top and back of the head (behind the ears).  These provide an indication of direct human touch.  Internal sensors include twin accelerometers, a temperature sensor and battery-level monitoring.

\subsection{Affect} \label{sec:EMO}

\emph{MiRo} represents its affective state (emotion, mood and temperament) as a point in a two-dimensional space covering `valence' (unpleasantness-pleasantness) and `arousal' (calm-excited) \cite{Collins2015b}.  Events arising in \emph{MiRo}'s sensorium are mapped into changes in affective state: for example, stroking \emph{MiRo} drives valence in a positive direction, whilst striking \emph{MiRo} on the head drives valence in a negative direction.  Baseline arousal is affected by general sound/light levels as well as the time of day; \emph{MiRo} is more active in the daytime.  An individual event can cause an acute change: for example, a very loud sound might raise arousal and decrease valence.  \emph{MiRo}'s movements are modulated by its affective state, and it also expresses itself using a set of `social pattern generators' that drive light displays, movement of the ears, tail, eyelids and - of particular relevance here - vocalisation.

\section{MiRo's Vocalisation System} \label{sec:VOC}

\subsection{Vocal Design Environment} \label{sec:VDE}

Prior to the development of \emph{MiRo}, the first author had already constructed a real-time parametric general-purpose mammalian vocal synthesiser (in accordance with the principles outlined in Section~\ref{sec:MAM}) aimed at designing `appropriate' vocalisations for a range of different animals and robots \cite{Moore2016}.  The design environment is implemented in `Pure Data' (referred to as ``Pd'') - an open-source visual dataflow programming language specifically created to operate with real-time audio\footnote{Pd (and its professional counterpart: MAX-MSP) is commonly used in music studios.} \cite{PD}.  The latest version is available for free download at \url{http://www.dcs.shef.ac.uk/~roger/downloads.html}.

The key Pd objects in the design software correspond to the \texttt{[lungs]}, [larynx], \texttt{[vocal tract]} and \texttt{[post-processing]}.  The command to vocalise initiates simulated airflow from the \texttt{[lungs]} object with an amplitude that is calculated from the flow rate.  The duration of the vocalisation is then calculated as a function of the flow rate and the lung capacity, and this is used to determine the period of the entire utterance.  

These signals and messages are passed to the \texttt{[larynx]} object which modulates the energy flow using the simulated action of one or two\footnote{It is well established that two excitation signals slightly offset in fundamental frequency give the resulting vocalisation a distinct robotic timbre.} sets of vocal folds vibrating at a fundamental frequency determined by the body mass, which is itself modulated by the utterance period.  With default settings, this gives rise to a rise-fall intonation pattern.  The voice quality, degree of aspiration (noise), level of quantisation and pitch difference between the two sets of vocal folds are all input parameters to the \texttt{[larynx]} object and influence the signal that is output to the \texttt{[vocal tract]} object.  

The \texttt{[vocal tract]} object simulates three acoustic resonances (formants) using band-pass filters whose frequencies are determined by the vocal tract length and the degree of mouth opening (using Equation~\ref{eq:FR}).  A syllabic rate parameter controls the opening and closing of the mouth.

\begin{figure}[ht]
        \centering
        \includegraphics[width=0.75\linewidth]{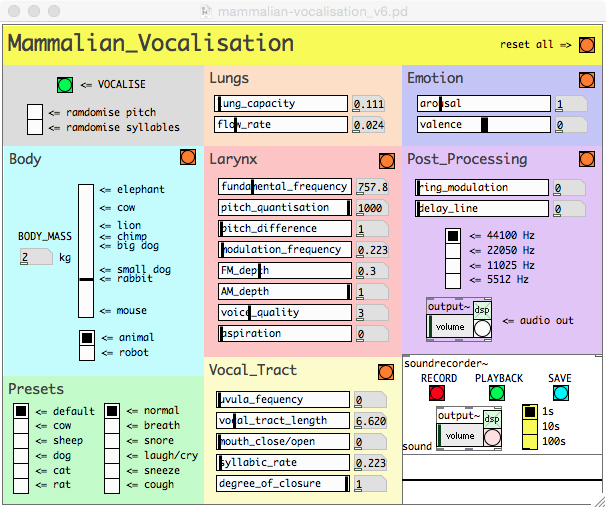}
        \caption{{\it Screenshot of the Pd GUI for the real-time parametric general-purpose mammalian vocal synthesiser that was used to design MiRo's voice.}}
        \label{fig:GUI}
      \end{figure}

Control parameters are set via a GUI using appropriate buttons and sliders (see Fig.~\ref{fig:GUI}).  This facilitates real-time adjustment of the vocalisation, and greatly enhances the process of designing different voices.  In principle, it is possible to set every parameter independently.  However, in practice, there are a number of potential dependencies (as described in Section~\ref{sec:MAM}).  As a result, setting the body size to a particular value also sets:
\begin{itemize}[noitemsep,topsep=0pt]
	\item the lung capacity (using Equation~\ref{eq:LC}),
	\item the breathing rate (using Equation~\ref{eq:BR}),
	\item the flow rate (using Equation~\ref{eq:FR2}),
	\item the fundamental frequency (using Equation~\ref{eq:FX}), and
	\item the vocal tract length (using Equation~\ref{eq:VTL}).
\end{itemize}

The software also provides a number of preset settings.  For example, it is possible to select particular animals (such as a rat, cat, dog, sheep, dog or cow in the current version), and also select different types of vocalisation (such as normal, breathing, snoring, laughing/crying, sneezing and coughing).  Selecting one of these presets simply moves all of the sliders to particular predetermined positions.  After selecting a preset it is still possible to vary any/all of the parameters as required in order to achieve a particular design objective.
      
\subsection{Implementation on MiRo} \label{sec:IMP}

The real-time parametric general-purpose mammalian vocal synthesiser design environment described in Section~\ref{sec:VDE} above was used ($a$) as a basis for implementing \emph{MiRo}'s biomimetic vocalisation system on the robot platform outlined in Section~\ref{sec:ARCH} and ($b$) to determine the appropriate parameter settings.  Accordingly, \emph{MiRo}'s synthesis software (programmed in C) was structured to simulate the flow of energy through a mammalian vocal apparatus with body mass corresponding to a land mammal of an equivalent size ($\sim$2 kg).  The vocalisation modules were integrated into \emph{MiRo}'s `biomimetic core' (corresponding to the `brainstem' in Fig.~\ref{fig:ARCH}).

The robot has a breathing rhythm ($\sim$0.7 Hz), the frequency of which is linked to arousal (see Section~\ref{sec:EMO}), and vocalisation is initiated stochastically during the exhalation phase.  Breathing is simulated as cyclic airflow into and out of the lungs with an amplitude and duration that is calculated from the flow rate, lung capacity and body mass.  When vocalising, the larynx modulates the airflow using the simulated action of a set of vocal folds vibrating at a fundamental frequency ($\sim$760 Hz) that is also determined by the body mass.  The vocal tract then simulates three formants using band-pass filters whose frequencies are determined by the vocal tract length ($\sim$6.6 cm) and the degree of mouth opening.  A syllabic rate parameter controls the opening and closing of the mouth, and a vibrating uvula adds a `cute' robotic timbre to the voice.  It was decided \emph{not} to employ two sets of vocal folds.

In order to allow the injection of emotion into the vocalisations, parameters were linked to \emph{MiRo}'s two-dimensional affect map (as discussed in Section~\ref{sec:EMO}).  Arousal modulates the rate of airflow and, thereby, the amplitude and tempo of the vocalisations; high arousal leads to high airflow and short vocalisations (and \emph{vice versa}).  Valence influences the variance of the fundamental frequency and the voice quality; high valence leads to expressive vocalisation whereas low valence produces more monotonic utterances.

\subsection{Example Vocalisations} \label{sec:EGS}

\begin{figure}[h!]
\begin{minipage}[b]{.5\linewidth}
\centering
\includegraphics[width=\linewidth]{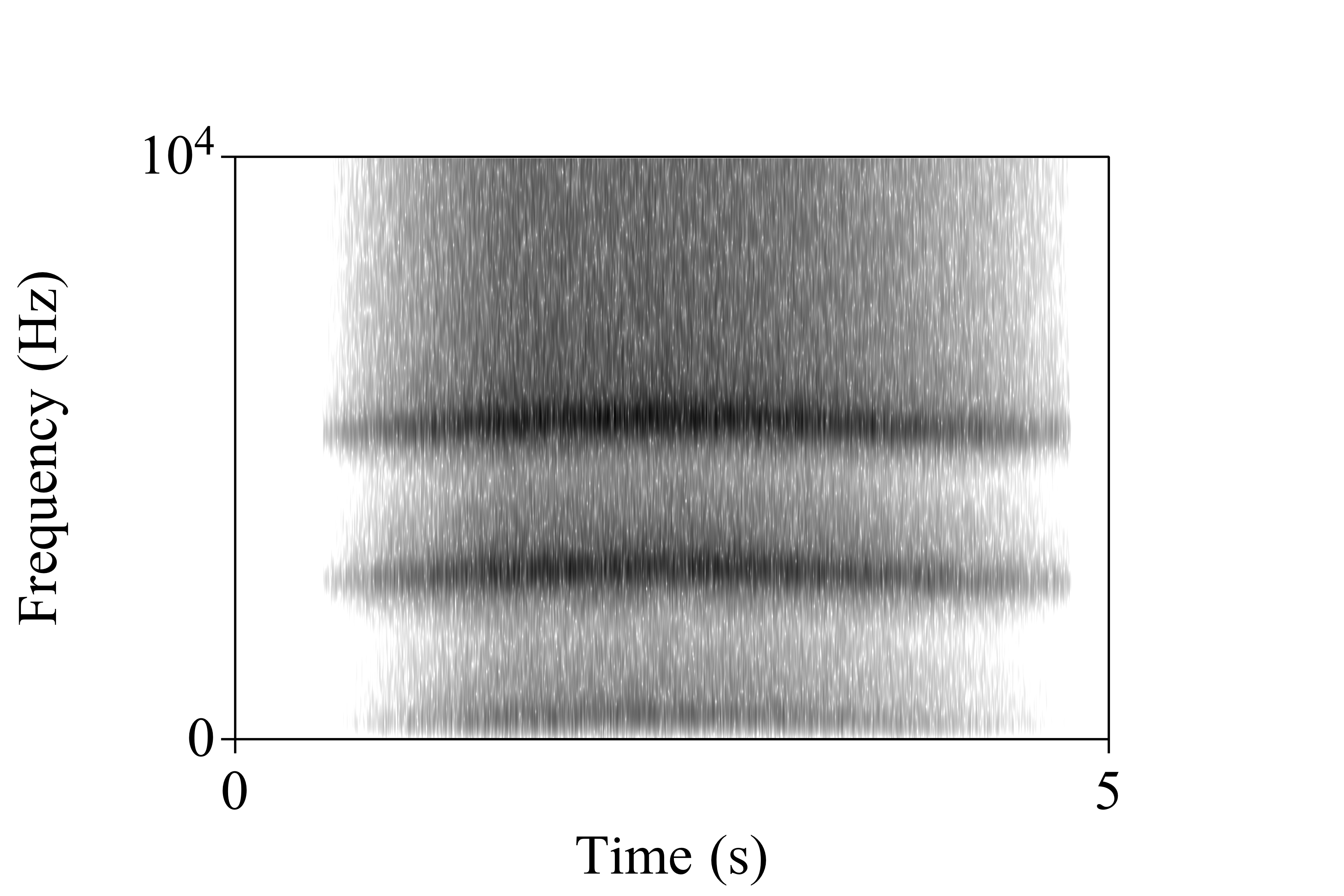}
Breath
\end{minipage}
\begin{minipage}[b]{.5\linewidth}
\centering
\includegraphics[width=\linewidth]{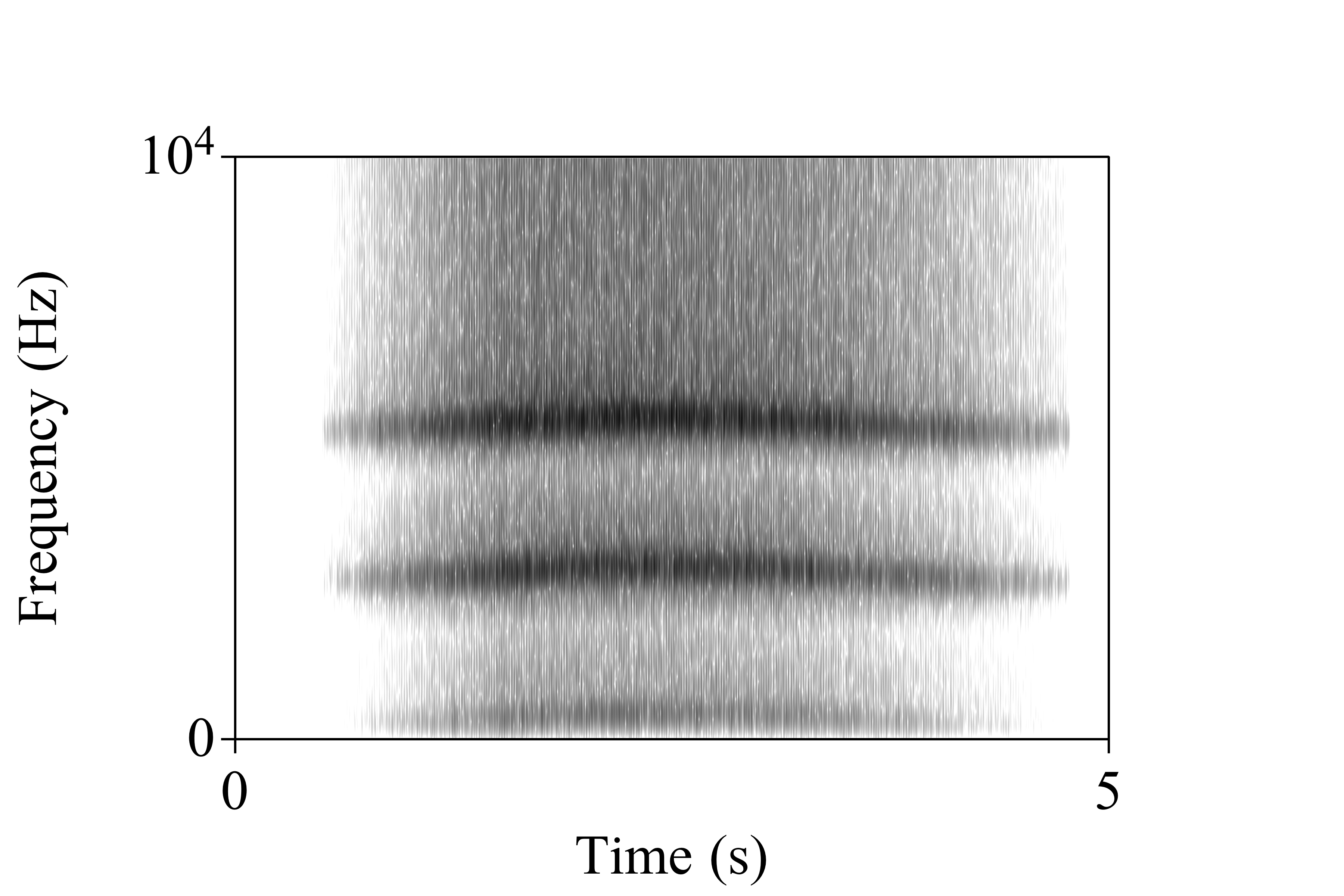}
Snore
\end{minipage}
\caption{Spectrograms of \emph{MiRo}'s vocalisations for an exhaled breath and an inhaled snore.  The dark bars indicate the concentration of energy at the formant resonances, and the vertical striations in the snore reflect the vibrating uvula.}
\label{fig:BS}
\end{figure}

As an example of the vocalisation system in operation, Fig.~\ref{fig:BS} shows spectrograms\footnote{A time-frequency energy plot commonly used to analyse speech and audio signals.} for two basic sounds - breathing and snoring.  These \emph{unvoiced} vocalisations are generated using noise as a excitation signal (as described in Section~\ref{sec:EXC}), and the spectrograms clearly illustrate the three-formant resonant structure of \emph{MiRo}'s simulated vocal tract (Section~\ref{sec:FLT}).  For these particular sounds, the vocal tract is fairly static throughout, hence there is little variation in the formant frequencies.

\begin{figure}[h!]
\begin{minipage}[b]{.5\linewidth}
\centering\includegraphics[width=\linewidth]{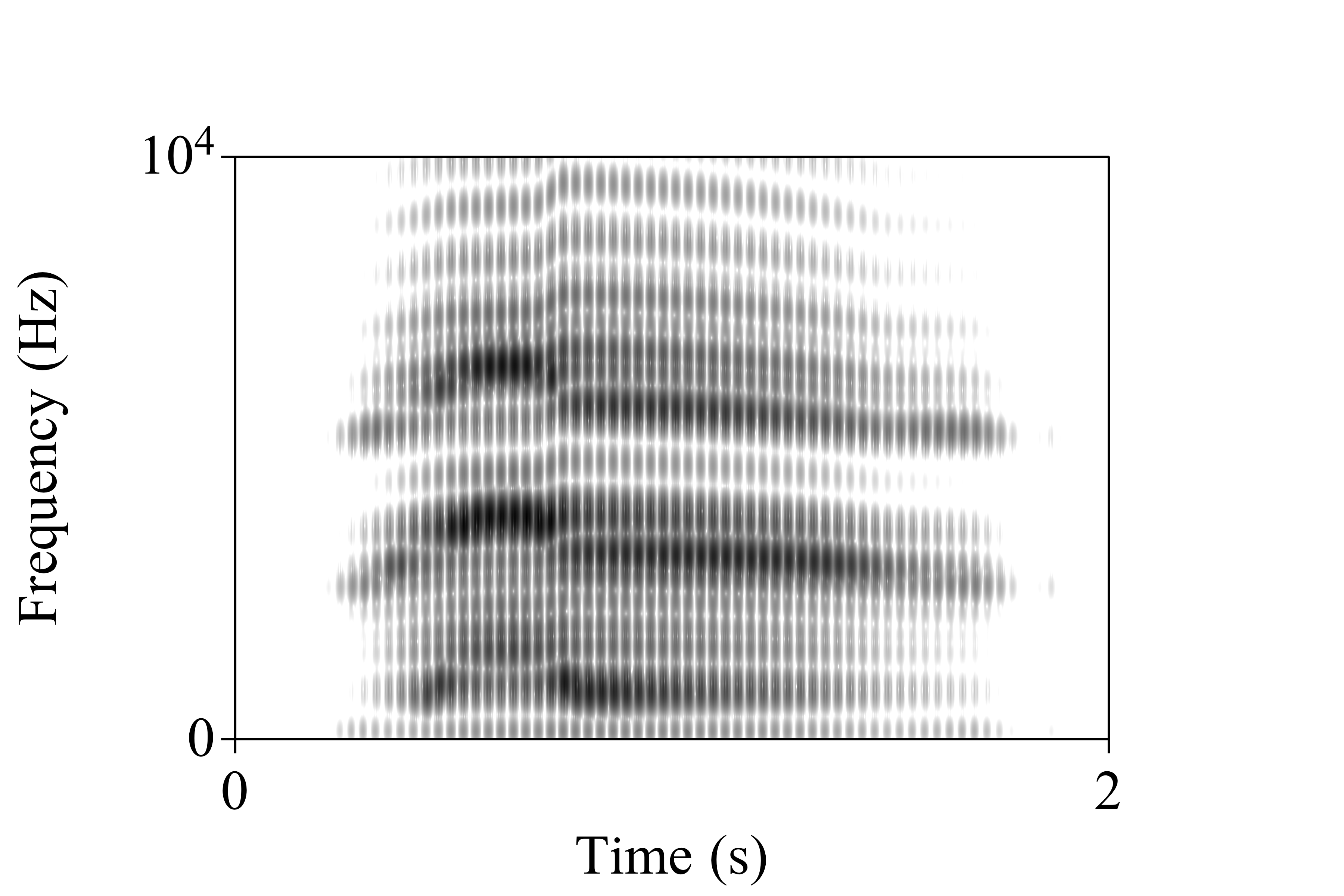}
$-ve$ arousal $-ve$ valence
\end{minipage}
\begin{minipage}[b]{.5\linewidth}
\centering
\includegraphics[width=\linewidth]{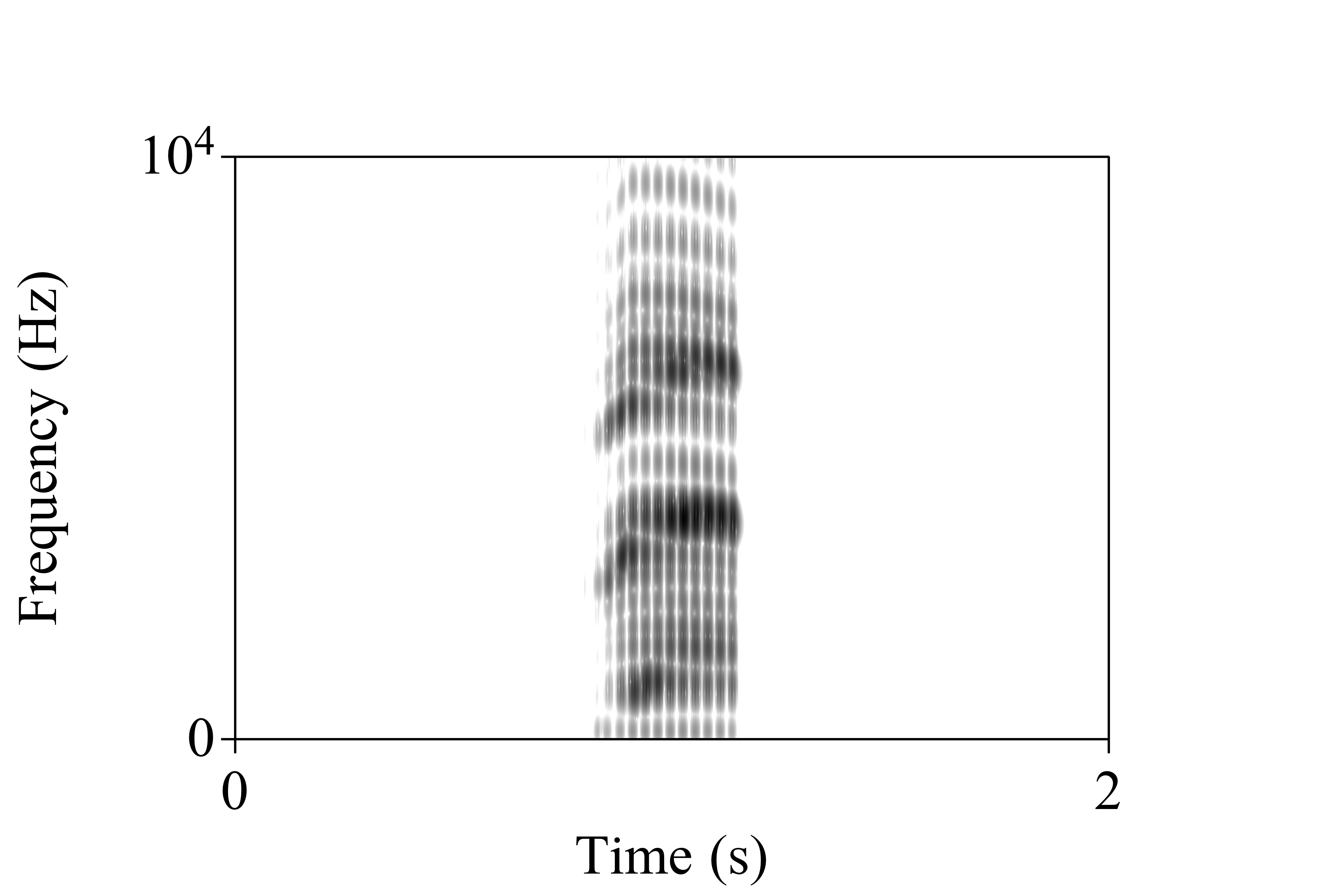}
$+ve$ arousal $-ve$ valence
\end{minipage}
\begin{minipage}[b]{.5\linewidth}
\centering
\includegraphics[width=\linewidth]{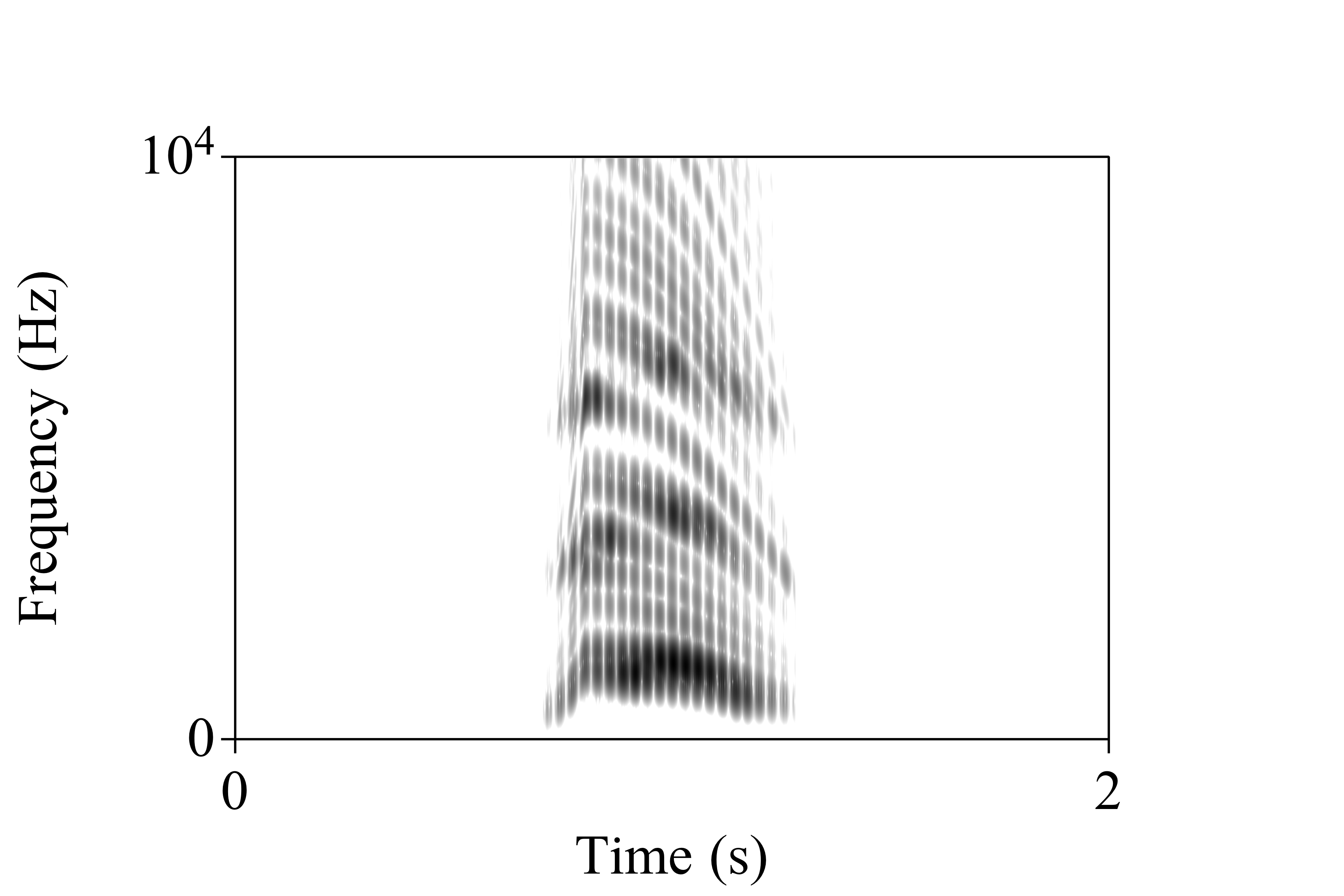}
$-ve$ arousal $+ve$ valence
\end{minipage}
\begin{minipage}[b]{.5\linewidth}
\centering\includegraphics[width=\linewidth]{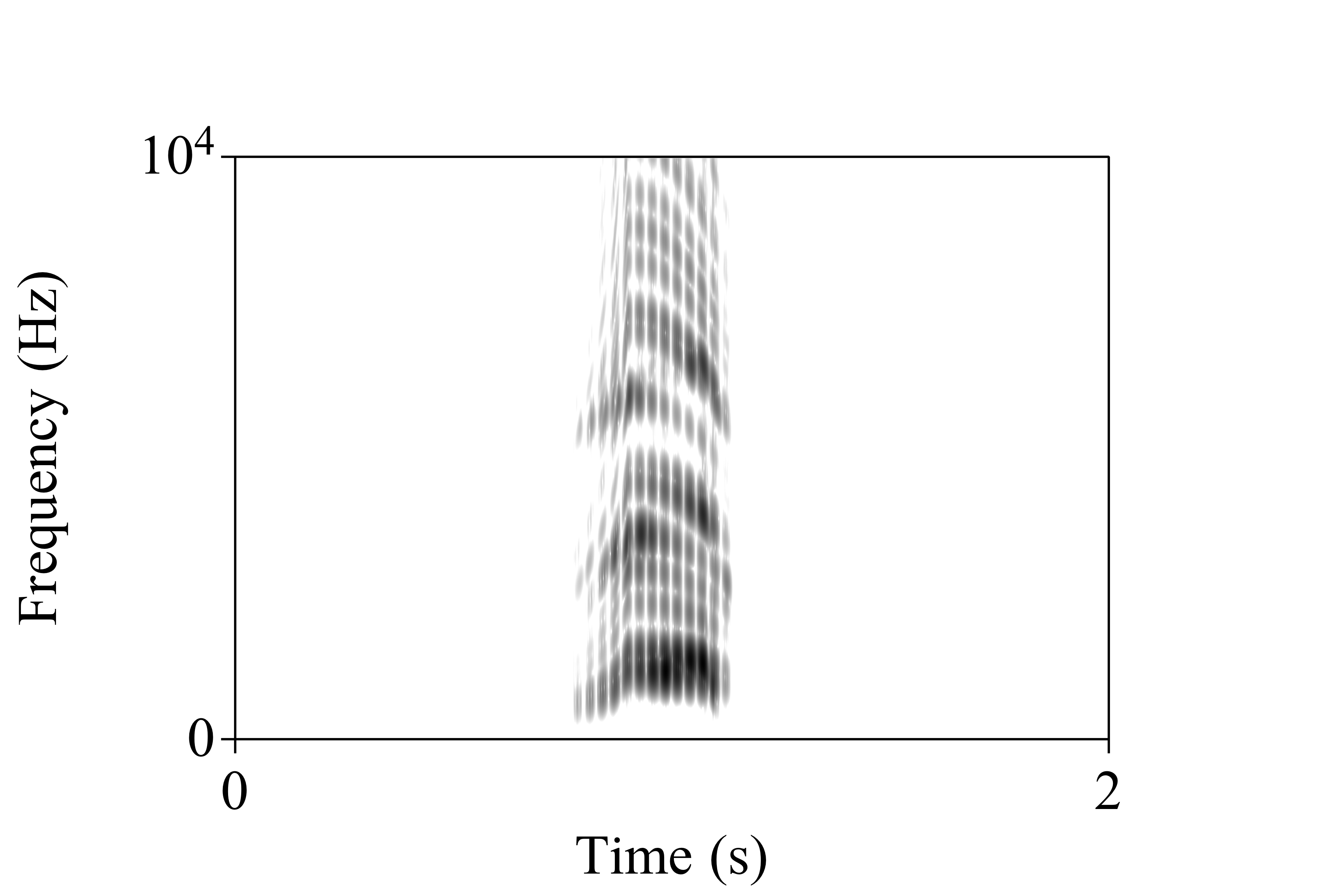}
$+ve$ arousal $+ve$ valence
\end{minipage}
\caption{Spectrograms of \emph{MiRo}'s vocalisations resulting from different values for arousal and valence.}
\label{fig:AV}
\end{figure}

In contrast, Fig.~\ref{fig:AV} shows spectrograms for \emph{voiced} vocalisations with different affective states (as described in Section~\ref{sec:EMO}).  As can be seen, these sounds are more dynamic than those shown in Fig.~\ref{fig:BS}, mainly due to the opening and closing of the mouth.  In addition, the formants vary more with positive valence (due to larger mouth opening), and the durations are shorter with high arousal (due to higher airflow).

\section{Summary and Conclusion} \label{sec:CONC}

This paper has described the design and implementation of \emph{MiRo}'s biomimetic vocalisation system.  Based on the principles underlying vocalisation in land mammals, it has been shown how the key characteristics of \emph{MiRo}'s vocalisations were selected using a real-time parametric general-purpose mammalian vocal synthesiser tailored to the specific physical characteristics of the robot.  It has been explained how these design decisions were ported onto \emph{MiRo}'s hardware/software platform and integrated into the robot's overall control architecture.  The novel outcome has been the creation of an `appropriate' voice for \emph{MiRo} that is perfectly aligned to the physical and behavioural affordances of the robot \cite{Moore2015}.  As such, it successfully avoids the `uncanny valley' effect caused by mismatched perceptual cues \cite{Mori1970,Moore2012} and contributes strongly to the effectiveness of \emph{MiRo} as an attractive interactive device.

\section{Acknowledgements}

This work was partially supported by the European Commission [grant numbers EU-FP6-507422, EU-FP6-034434, EU-FP7-231868 and EU-FP7-611971], and the UK Engineering and Physical Sciences Research Council (EPSRC) [grant number EP/I013512/1].

\end{document}